\documentclass{aa501}
\usepackage{epsfig}
\begin{document}

\title{The distance scale and Eddington efficiency of luminous quasars}

\author{P. Teerikorpi}
\institute{Tuorla Observatory,
University of Turku, FIN-21500 Piikki\"o, Finland}

\date{Received / Accepted}

\abstract{The relation between the central mass and quasar luminosity
($M_{\mathrm{BH}} \propto L^{\alpha}${\small \it FHWM}$^2$) links a given
Eddington ratio with a value of $H_0$, within a cosmology with fixed
($\Omega_\mathrm{m}$,$\Omega_{\Lambda}$). 
We point out that because the relation is calibrated at low $z$ using
distance independent
reverberation mapping to get the BLR size, the derived $M_{\mathrm{BH}}$
interestingly does not depend on $H_0$, while  $L/L_{\mathrm{Edd}}$ is
sensitive to $H_0$, but rather robust to changes of
$\Omega_{\Lambda}$ in the standard flat model.
This means, e.g., that enough of extragalactic objects
radiating at the Eddington limit could be used to study the global
Hubble constant in a new way, bypassing the local distance ladder.
The method could become practical when systematic errors in derived
$M_{\mathrm{BH}}$ are understood and objects with $L \la L_{\mathrm{Edd}}$
can be independently identified. As an illustration,
if we take a sample of tranquil very luminous quasars in the redshift
range $0.5 < z < 1.6$, and assume that they are radiating with
$L_{\mathrm{bol}}\leq
L_{\mathrm{Edd}}$, then the usual numerical factors used for calculating
$M_{\mathrm{BH}}$ and $L_{\mathrm{bol}}$ would lead to the result that
the Hubble constant must be larger than 45 km/s/Mpc.
}
\titlerunning{Distance scale and Eddington efficiency}

\maketitle

\section{Introduction}

Recent years have made it possible to infer masses of the
compact nuclei in galaxies and quasars. Important relations
were found between this mass and the host galaxy's  mass
(Ferrarese \& Merritt 2000; Gebhardt et al. 2000).
The mass $M_{\mathrm{BH}}$ has been determined by primary methods
for relatively nearby objects (such as reverberation
mapping giving a size of the broad line region $R_{\mathrm{BLR}}$) and by
secondary methods for more distant AGN's
(such as the relation between $R_{\mathrm{BLR}}$ and optical luminosity;
see Vestergaard 2004). The masses for quasars come
mostly from the $L_{\mathrm{opt}}$ --
$R_{\mathrm{BLR}}$ relation, with
a velocity parameter given by the emission line width:
\begin{equation}
M_{BH} = a  [\lambda L_{\lambda}(5100Å)/10^{44}
\mathrm{erg\, s}^{-1}]^{\alpha}
\mathrm{{\small \it FHWM}}^2
\label{BHmass}
\end{equation}
The exponent $\alpha$ has
been given values
ranging from $0.5$ to $0.7$ (Netzer 2003).

Here we point out that the way the BLR size vs.
luminosity relation is calibrated, has an interesting implication when
deriving BH masses and Eddington ratios within a Friedmann model,
and one may view radiators at the Eddington limit as possible
high-$z$ building
blocks of the distance scale.
Though there are still problems with systematic errors and now there is
no way to extract such radiators
independently, one should be aware of this prospect.

\section{Calibration and Eddington ratio}

The relation between the size $R_{\mathrm{BLR}}$
and luminosity $L$ is calibrated at low redshifts ($\la 0.2$) 
using an assumed value of $H_0$ (Kaspi et al. 2000).
\emph{It is important to note that as the size is obtained from a
light travel
time argument (``reverberation mapping''), it does not depend on the
distance scale} and the $R_{\mathrm{BLR}}$ vs. $L$ relation from the
calibrator sample just shifts along
the luminosity axis by a factor of $h^{-2} = (H/H_0)^{-2}$.
As the inferred luminosity of a higher-$z$ sample quasar is also changed by
this same factor, a change of $H_0$ does not change
BH masses at all,
but it does change  Eddington ratios $L/L_{\mathrm{Edd}}$ by a factor
of $h^{-2}$:
\begin{equation}
 L/L_{\mathrm{Edd}} = \frac{4\pi r_{\mathrm{L}}^2 f_{\mathrm{bol}}}
 {b M_{\mathrm{BH}}} 
\propto h^{-2}
\label{hdepend}
\end{equation}
We remind that $L_{\mathrm{Edd}} = 1.26
\times 10^{38}(M_{\mathrm{BH}}/M_{\sun})$ergs$^{-1}$, so the individual
values of $L_{\mathrm{Edd}}$ do not depend on
$H_0$.

It is this sensitivity to $H_0$ which makes Eddington radiators
interesting for the distance scale.   
If the size in the calibration were calculated, as usually, from an angular
quantity, then  $L/L_{\mathrm{Edd}}$ would change as $h^{-1}$.

Another important point is that changing the
other cosmological parameters ($\Omega_{\mathrm{m}}, \Omega_{\Lambda}$)
has a smaller influence on the Eddington ratio.
Then comoving distances change, depending on $z$, and one must
make a further individual correction to $L$ ($L_c = (r_2/r_1)^2 L$),
which affects  the Eddington ratio only as
$(r_2/r_1)^{2(1-\alpha)}$ (for $\alpha = 0.6$ this becomes
$(r_2/r_1)^{0.8}$). A similar tiny effect on
the calibration taking place at low redshifts may be generally ignored.

\section{The distance scale}

If one has reasons to think that some quasars radiate at $L_{\mathrm{Edd}}$,
one may infer which luminosity distance and, hence, which value of $H_0$
leads to this efficiency (for adopted $\Omega_m, \Omega_{\Lambda}$).
This could be a nice example of a definition once given:
a distance indicator is a method where a galaxy is placed in 3-D space so
that its observed properties agree with what we know about
galaxies, their constituents and the propagation of light (Teerikorpi 1997).

The method would have some positive sides for ideal
Eddington radiators:
(1) independence of the local distance ladder,
(2) probes the global distance scale,
(3) not influenced by the usual Malmquist bias,
(4) sensitive indicator of $H_0$, and
(5) rather robust to changes in $\Omega_{\Lambda}$
 in the $\Lambda$-dominated flat model.

The last two items relate, respectively, to the low-$z$ calibration
of the BLR size vs. luminosity relation using the
distance-scale independent reverberation mapping and to the favourable
exponent $0.5 \le \alpha < 1$ in the relation.

The Malmquist bias usually accompanies determination of distances
(e.g. Teerikorpi 1997). Here the usual
bias is absent: the calibration and derivation of the BLR size is made against
luminosity, hence a given $\log L$ predicts an unbiased size, hence
unbiased $M_{\mathrm{BH}}$ and  $L_{\mathrm{Edd}}$.  
This resembles the case of the inverse Tully-Fisher relation (Teerikorpi 1984).

At intermediate redshifts a small fraction of quasars
seem to radiate around $L_{\mathrm{Edd}}$,  while at $z < 0.5$ quasars
work at lower accretion rate (Woo \& Urry 2002; McLure \& Dunlop 2004).
However, there is presently no sure way to tell -- without the measured mass
-- if a quasar really radiates at $L_{\mathrm{Edd}}$. Together with
possible systematic errors, this forms an obstacle for practical
applications of the method.

\section{Systematic errors: a challenge}

Even if one could identify, say from a spectral property,
an object or a whole class radiating at the Eddington limit,
there are still systematic errors. Such may be caused by
the mass estimate $M_{\mathrm{BH}}$, involving the
BLR size versus luminosity relation, the exponent $\alpha$,
the factor $f$ in
the velocity parameter  as $f \mathrm{{\small \it FWHM}}$,
 and the estimate of $L_{\mathrm{bol}}$.

The factor $f$ that transforms the Doppler-broadened
emission line width {\small \it FWHM} into a relevant orbital velocity
is still discussed (Krolik 2001;Vestergaard 2002; Netzer 2003).
It affects the derived  $M_{\mathrm{BH}}$ as $f^2$.   For an isotropic
distribution $f=\sqrt{3}/2 \approx 0.87$ (Netzer 1990),
while McLure \& Dunlop (2001) suggest a larger factor $f =3/2$, from
a combination of isotropic and disk components. In McClure \& Dunlop (2002)
they showed modelling the FWHM distribution for an AGN sample that $f$
appears
to depend on the H$\beta$ emission line width (inclination dependence),
and for widths $>2800$
km/s (which formed 83 per cent of their sample) $f$ was rather close
to the isotropic value $f=\sqrt{3}/2$. This model got some support from
a sample of 10 Seyfert galaxies for which stellar velocity dispersion
measurements were available.  McLure \& Dunlop (2004) use $f=1$. 

We note that if the average value of $f$ were now much in error, say by
a factor of 2, one would expect a visible shift in the   $\log
 M_{\mathrm{BH}}$ vs. $M_R(\mathrm{bulge})$ relation for a sample
with mass estimates from $H\beta$ line widths and a sample with
actual dynamical estimates. However, a good agreement appears
in the study by McLure \& Dunlop (2002).

Going from  $M_{\mathrm{BH}}$ to  $\log L/L_{\mathrm{Edd}}$, one needs
the bolometric luminosity $L_{\mathrm{bol}}$. It is usually calculated using
 a constant correction $\log k$ to $\log \lambda L_{\lambda}(5100)$, equal
to about $\log 9.5$. There are larger variations from-quasar-to-quasar
in this bolometric correction (Elvis et al. 1994).
However, one may hope that for a uniform AGN class these variations
may be lower, and in any case, here it is the average value of the correction
$k$ and its error that are important.
 Judging from recent studies a systematic error, caused
by an incorrect average continuum, is narrowing to within
$\pm 0.2$ (Kaspi et al. 2000; Vestergaard 2004; Warner et al. 2004).

Even if systematic errors were in control, a large enough sample
of Eddington radiators is needed to give useful limits to $H_0$.
If for a single object one may optimistically
expect a future $1\sigma$ accuracy of 0.3 in $\log L/L_{\mathrm{Edd}}$,
this transforms into 0.15 in $\log H_0$. E.g., 25 Eddington
radiators would
thus fix $\log H_0$ within $\pm 0.03$ ($1\sigma$), forgetting the small
uncertainty due to the $\Omega$ parameters of the Friedmann model.
With $h_0 \approx 0.6$, this would imply $1 \sigma$ error bars of
about $\pm 5$ km/s/Mpc. 

If the objects actually radiate below $L_{\mathrm{Edd}}$, then
the inferred $H_0$ is a kind of lower limit. Hence, even the usually
valid assumption that $L \leq L_{\mathrm{Edd}}$ may have interesting
bearing on the distance scale.

\section{A simple illustration}

Having this in mind, it would be interesting to have a sample of
quasars, probably at close to or below the Eddington limit.
To illustrate we consider a class ``AI'' of luminous quasars,
first proposed to exist 
without any consideration of $L_{\mathrm{Edd}}$, and
radiating around $M_{\mathrm{min}} \approx -26.0 + 5 \log h_{100}$,
a minimum brightness V magnitude (Teerikorpi 2000).

These quasars are found at redshifts
$0.5 \div 1.7$, when according to McLure \& Dunlop (2004) in a
sample of the first data release SDSS quasars in the interval
$0.5 \la z \la 2$ ``the Eddington luminosity is still a relevant
physical limit to the accretion rate of luminous quasars'', but
below $z \approx 0.5$ the efficiency is lower.
These are perhaps optically the most luminous and at the same time tranquil
quasars, optically rather inactive.
They probably do not contain a strong beamed optical component.
The limit $M_{min} < -25.5$ accounts for the fact that fainter quasars
are more violent optically as measured both with optical variability
and polarization (Teerikorpi 2000). 

\begin{table}
\caption[]{Data ($M_{min} < -25.5$ mag, $0.5 < z < 1.6$)}
\begin{tabular}{lcccccr} 
 RA & $\delta$ & $z$ & $M_{min}$ & $\log L_{bol}$ & $\log M_{BH}$ & ref \\ \hline \hline 
0024 & +2225 & 1.118 & -26.0 & 47.40 & 9.45 & 2 \\
0405 & -1219 & 0.574 & -26.0 & 47.69 & 9.58 & 1 \\
0414 & -0601 & 0.781 & -25.7 & 46.97 & 9.52 & 2 \\
0454 & +0356 & 1.345 & -26.7 & 47.60 & 9.39 & 2  \\
%0538 & +4949 & 0.545 & -25.0 & 46.71 & 9.69 & 1  \\
0637 & -7513 & 0.651 & -26.3 & 47.46 & 9.54 & 1  \\
%0738 & +3119 & 0.630 & -25.2 & 47.24 & 9.52 & 1 \\
%0809 & +4822 & 0.871 & -25.0 & 46.89 & 8.12 & 1  \\
0952 & +1757 & 1.472 & -25.9 & 47.41 & 9.23 & 2    \\
%1136 & -1334 & 0.554 & -25.2 & 47.06 & 8.89 & 1    \\
%1137 & +6604 & 0.646 & -25.0 & 47.15 & 9.48 & 1 \\
%1340 & +2859 & 0.905 & -25.3 & 46.89 & 9.88 & 2     \\
%1354 & +1933 & 0.720 & -25.1 & 47.43 & 9.57 & 1     \\
1458 & +7152 & 0.905 & -25.7 & 47.28 & 9.14 & 1 \\
%1656 & +0519 & 0.879 & -25.1 & 47.56 & 9.78 & 1  \\
%1828 & +4842 & 0.692 & -24.8 & 47.09 & 9.98 & 1  \\
1954 & -3853 & 0.626 & -25.8 & 46.61 & 8.75 & 1  \\
%2128 & -1220 & 0.501 & -25.0 & 47.03 & 9.71 & 1  \\
%2143 & -1539 & 0.700 & -25.3 & 46.96 & 7.81 & 1 \\
2216 & -0350 & 0.901 & -26.0 & 47.52 & 9.40 & 1 \\
%2251 & +1552 & 0.859 & -25.3 & 47.61 & 9.32 & 1 \\
2255 & -2814 & 0.926 & -26.1 & 47.32 & 9.32 & 1  \\
2344 & +0914 & 0.677 & -25.8 & 47.38 & 9.44 & 1 \\ \hline
\end{tabular}
\noindent \underline{References:} 1. Woo \& Urry (2002) 2. Shields et al. (2003)
\end{table}

About 30 potential AI objects with $z$ in the range $0.5 \div 1.7$
are found in our list of radio quasars with UBV photometry (Teerikorpi 2000).
Of these 11 have an entry in the compilations of $M_{\mathrm{BH}}$
(Woo \& Urry 2002; Shields et al. 2003),
when one limits the sample to  $M_{\mathrm{min}}< -25.6$.
Those authors use the exponents 0.7 and 0.5, respectively, in
Eq.(\ref{BHmass}). Here we have only made the adjustment
to the adopted cosmology, and then the bulk of these objects were found to
concentrate not far from a ($M_{\mathrm{BH}}$, $L_{\mathrm{bol}}$)
line corresponding to a constant Eddington ratio.

We take for the $L_{\mathrm{bol}}$ vs.
$M_{\mathrm{BH}}$ diagram in Fig.1 9 objects with highest
Eddington ratios,
\footnote{Two quasars in Table 1 would lie well below the
other ones in Fig.1: 0414-0601 and 1954-3853. The latter one has
optical polarization of $11$ percent and variability at least
$0.8$ mag (Table 1 in P2), hence it is optically active. For these objects
the Eddington ratio is much less than the ratio for those displayed
in Fig.1.}
and show the effect of $H_0$, for the standard model
($\Omega_\mathrm{m},
\Omega_{\Lambda}$) = (0.3,0.7). With $h_{100} = 0.45$ and 0.80,
these AIs appear to radiate above and below the Eddington value,
respectively. There is just a vertical shift by the factor $2\log{80/45}$
leaving the masses $M_{\mathrm{BH}}$ the same (sect.2).
Such limits on $H_0$ would be excluded if the AIs were
{\it known} to radiate at $L_{\mathrm{Edd}}$ and if there were
no systematic errors (sect.4). The plausible assumption
$L_{\mathrm{bol}}\leq L_{\mathrm{Edd}}$ makes $h_{100} = 0.45$ a strict
lower limit.  

\begin{figure}
\epsfig{file=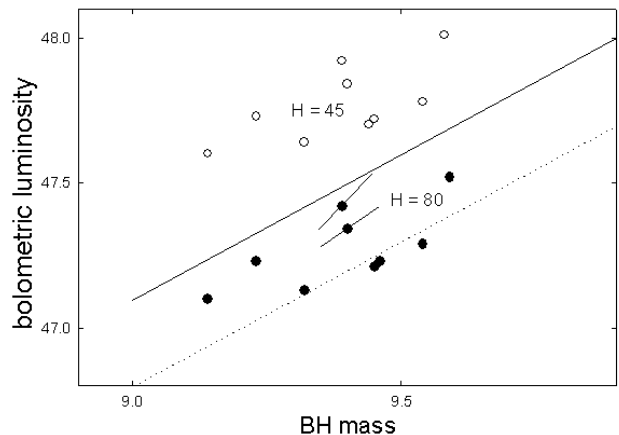, angle=0, width=9cm}
\caption{A sample of luminous AI quasars in the $z$ interval $0.6 \div 1.5$
 in the $\log L_{\mathrm{bol}}$ vs. $\log M_{\mathrm{BH}}$
diagram for two values of $H_0$ in the world model
$(\Omega_\mathrm{m},\Omega_{\Lambda})$ = (0.3,0.7).
The upper line indicates
the Eddington luminosity, the lower line is $0.5L_{\mathrm{Edd}}$.
A change of $H_0$
causes a vertical shift. For two quasars we show how
$\Delta \Omega_{\Lambda} = \pm 0.15$ in the flat model
influences their positions. If these quasars are radiating at $L_{Edd}$
or less, then for the numerical factors used for calculating
$L_{bol}$ and $M_{BH}$, 45 kms$^{-1}$/Mpc is a strict lower
limit for the Hubble constant.}
\label{fig2}
\end{figure}

If we change $\Omega_{\Lambda}$ by $\pm 0.15$ (keeping 
$\Omega_{\mathrm{tot}} =1$), taking the $\sim 2\sigma$ error bars from
SNIa's (Tonry et al. 2003; Knop et al. 2003),
one sees just a small dependence of the Eddington ratio on
$\Delta \Omega_{\Lambda}$ in Fig.1, where for two quasars
are indicated the position shifts for $\Delta \Omega_{\Lambda} =
\pm 0.15$. The steeper slope corresponds to $\alpha = 0.5$, the shallower
one to $\alpha = 0.7$.

\section{Concluding remarks}

Though presently this approach does not add much independent
information about the distance scale, it is noteworthy that
the natural assumption that
$L \leq L_{Edd}$ for the considered quasars leads to a lower limit
$h > 0.5$ when the usual values of the parameters are used in the calculation
of the BH mass and bolometric luminosity. Hence at least
a consistent picture emerges. 

There is still room for high-$z$ methods
to study $H_0$ bypassing the local distance ladder, complementing
those based on the Sunuyaev-Zeldovich effect
% (Silk \& White 1978)
and the time delay in gravitational lens images.
% (Refsdal 1964).
Furthermore, there are still problems with the local Cepheid-based
distance scale (Paturel \& Teerikorpi 2005).
In the present paper we have shown how the study of the Eddington
efficiency
is also interestingly connected with the problem of the global
distance scale, when in the range
$0.5 \la z \la 2$ ``the Eddington luminosity is still a relevant
physical limit to the accretion rate of luminous quasars'' (McLure \&
Dunlop 2004).

\section*{Acknowledgements}
This study has been supported by the Academy of Finland
("Fundamental questions of observational cosmology").

{}


\begin{thebibliography}{}
%\bibitem[2003]{bettoni03}
%Bettoni, D., Falomo, R., Fasano, G, Govoni, F., 2003, A\&A 399, 869
%%\bibitem[2001]{chelouche01}
%%Chelouche, D. \& Netzer, H., 2001, MNRAS 326, 916
%\bibitem{blanchard03}
%Blanchard A., Douspis M., Rowan-Robinson M., Sarkar S., 2003, A\&A 412, 35
\bibitem{elvis94}
Elvis M., Wilkes B.J., McDowell J.C. et al., 1994, ApJSS 95, 1
%\bibitem{fabian90}
%Fabian, A.C.,  Crawford, C.S.: 1990, MNRAS~247, 439
\bibitem{ferrarese00}
Ferrarese L.,  Merritt D., 2000, ApJ~539, L9
%%\bibitem[2002]{ferrarese02}
%%Ferrarese, L., 2002, ApJ 578, 90
\bibitem{gebhardt00}
Gebhardt K., Bender R., Bower G. et al., 2000, ApJ~539, L13 
\bibitem{kaspi00}
Kaspi S., Smith P.S., Netzer H., Maoz D., Jannuzi B.T.,
Giveon U., 2000, ApJ~533, 631
\bibitem{knop03}
Knop, R.A., Adering G., Amanullah R., et al., 2003, ApJ 598, 102
%\bibitem{kormendy95}
%Kormendy, J.,  Richstone, D.: 1995, ARA\&A 33, 581
\bibitem{krolik01}
Krolik J.H., 2001, ApJ 551, 72
%\bibitem{magorrian98}
%Magorrian, J., Tremaine, S., Richstone, D. et al.: 1998, AJ 115, 2285
\bibitem{mclure01}
McLure R.J.,  Dunlop J.S., 2001, MNRAS 327, 199
\bibitem{mclure02}
McLure R.J., Dunlop J.S., 2002, MNRAS 331, 795
\bibitem{mclure03}
McLure R.J., Dunlop J.S., 2004, MNRAS 352, 1390
\bibitem{netzer90}
Netzer, H., 1990, in Active Galactic Nuclei, ed. T.J.-L. Courvoisier
\& M. Mayor (Berlin: Springer), 137
\bibitem{netzer03}
Netzer H., 2003, ApJ 583, L5
%\bibitem{nilsson03}
%Nilsson, K., Pursimo, T., Heidt, J., Takalo, L.O., Sillanp\"{a}\"{a}, A.,
%Brinkmann, W., 2003, A\&A 400, 95
%%\bibitem[1996]{odea96}
%%O'Dea, C.P., Stanghellini, C., Baum, S., Charlot, S., 1996, ApJ 470, 806
%\bibitem{paturel04}
%Paturel, G., Teerikorpi, P.: 2004, A\&A 413, L31
\bibitem{paturel05}
Paturel G., Teerikorpi P., 2005, A\&A (in press)
%%\bibitem[1999]{pursimo99}
%%Pursimo, T., Nilsson, K., Teerikorpi, P. et al., 1999, A\&ASS 134, 505   
%\bibitem{refsdal64}
%Refsdal S., 1964, MNRAS 128, 307
\bibitem{shields03}
Shields G.A., Gebhardt K., Salviander S. et al., 2003, ApJ 583, 124
%\bibitem{silk78}
%Silk J., White S.D.M., 1978, ApJ 226, L103
%\bibitem{spergel03}
%Spergel D.N., Verde L., Peiris H.V. et al., 2003, ApJS 148, 175 
%\bibitem{sun89}
%Sun, W-H., Malkan, M.: 1989, ApJ 346, 68
\bibitem{teerikorpi81}
Teerikorpi P., 1984, A\&A 141, 407
\bibitem{teerikorpi97}
Teerikorpi P., 1997, ARA\&A 35, 101
\bibitem{teerikorpi00}
Teerikorpi P., 2000, A\&A 353, 77
%\bibitem{teerikorpi01}
%Teerikorpi, P.: 2001, A\&A 375, 752 (P3)
%\bibitem{teerikorpi02}
%Teerikorpi, P., Paturel, G., 2002, A\&A 381, L37
\bibitem{tonry03}
Tonry J.L., Schmidt B.P., Barris B. et al., 2003, ApJ 594,1
%%\bibitem[2003]{willot03}
%%Willot, C.J., McClure, R.J. \& Jarvis, M.J., 2003, ApJ 587, L15
\bibitem{vestergaard02}
Vestergaard M., 2002, ApJ 571, 733
\bibitem{vestergaard04}
Vestergaard M., 2004, ApJ 601, 676
%% proceedings of Multiwavelength AGN Surveys, ed. R. Mujica and R. Maiolino
%% [astro-ph/0407325]
\bibitem{warner04}
Warner C., Hamann F., Dietrich M., 2004, ApJ 608, 136
\bibitem{woo02}
Woo J.-H., Urry C.M., 2002, ApJ 579, 530
\end{thebibliography}
\end{document}